\documentclass[aps,prl,groupedaddress,fleqn,twocolumn,10pt]{revtex4}
\pdfoutput=1
\usepackage{graphicx}
\usepackage{amsmath}
\usepackage{gensymb}
\usepackage{float}
\usepackage{physics}
\usepackage{xcolor}

\begin{document}

\raggedbottom

\title{Minimal quantum viscosity from fundamental physical constants}
\author{K. Trachenko}
\affiliation{School of Physics and Astronomy, Queen Mary University of London, Mile End Road, London, E1 4NS, UK}
\author{V. V. Brazhkin}
\affiliation{Institute for High Pressure Physics, RAS, 108840, Troitsk, Moscow, Russia}

\begin{abstract}
Viscosity of fluids is strongly system-dependent, varies across many orders of magnitude and depends on molecular interactions and structure in a complex way not amenable to first-principles theories. Despite the variations and theoretical difficulties, we find a new quantity setting the minimal kinematic viscosity of fluids: $\nu_m=\frac{1}{4\pi}\frac{\hbar}{\sqrt{m_em}}$, where $m_e$ and $m$ are electron and molecule masses. We subsequently introduce a new property, the ``elementary'' viscosity $\iota$ with the lower bound set by fundamental physical constants and notably involving the proton-to-electron mass ratio: $\iota_m=\frac{\hbar}{4\pi}\left({\frac{m_p}{m_e}}\right)^{\frac{1}{2}}$, where $m_p$ is the proton mass. We discuss the connection of our result to the bound found by Kovtun, Son and Starinets in strongly-interacting field theories.
\end{abstract}

\maketitle

\section{Introduction}

Several important physical properties can be expressed in terms of fundamental physical constants including, for example, the Bohr radius, the Rydberg energy and fine structure constant \cite{ashcroft,barrow}. These properties serve as a basis for atomic units and importantly set the scale of energy and length: the Bohr radius gives a characteristic interatomic distance in condensed matter phases on the order of Angstroms and the Rydberg energy gives a characteristic binding energy on the order of several eV. It is interesting to see whether a {\it transport} property such as viscosity or diffusion can be similarly expressed in terms of fundamental constants, setting their characteristic scale. Here, we find a quantum quantity setting the minimal kinematic viscosity of fluids, $\nu_m$, as

\begin{equation}
\nu_m=\frac{1}{4\pi}\frac{\hbar}{\sqrt{m_em}}
\label{nu0}
\end{equation}

\noindent where $m_e$ is the electron mass and $m$ is the mass of the molecule set by the nucleon mass. For atomic hydrogen with the mass given by the proton mass $m_p$, $\nu_m$ is defined by fundamental constants only.

We subsequently introduce a new property: the ``elementary'' viscosity $\iota=\nu_m m$ with the lower bound $\iota_m$ set by fundamental physical constants as

\begin{equation}
\iota_m=\frac{\hbar}{4\pi}\left({\frac{m_p}{m_e}}\right)^{\frac{1}{2}}
\label{iota0}
\end{equation}

\noindent which is on the order of $\hbar$.

(\ref{iota0}) interestingly involves the proton-to-electron mass ratio, one of few dimensionless combinations of fundamental constants of importance in a variety of areas, including formation of stars, ordered molecular structures and life-supporting environment \cite{barrow}.

We recall that viscosity of fluids, $\eta$, varies in a wide range, from about 10$^{-6}$ Pa$\cdot$s for the normal component of He to 10$^{13}$ Pa$\cdot$s in viscous liquids approaching liquid-glass transition. $\eta$ strongly depends on temperature and pressure. $\eta$ is additionally strongly system-dependent and is governed by the activation energy barrier for molecular rearrangements, $U$, which in turn is related to the inter-molecular interactions and structure. This relationship in complicated in general, and no universal way to predict $U$ and $\eta$ from first principles exists (tractable theoretical models describe the dilute gas limit of fluids where perturbation theory applies, but not dense liquids of interest here \cite{chapman}). This is appreciated outside the realm of condensed matter physics: the difficulty of calculating the viscosity of water was compared to the problem of calculating the fundamental constants themselves \cite{weinberg}. As far as thermodynamic properties of liquids are concerned, the absence of a small parameter due to the combination of strong interactions and the absence of small oscillations is considered to disallow a possibility of calculating liquid thermodynamic properties in general form \cite{landau}. For example, theoretical calculation and understanding liquid energy and heat capacity has remained a long-standing problem \cite{granato} which started to lift only recently when new understanding of phonons in liquids came in \cite{f3}. In view of these problems related to liquid theory, the existence of universal $\nu_m$ (\ref{nu0}) and $\iota_m$ (\ref{iota0}) is notable.

We note that viscosity is mostly considered as a classical quantity. At the same time, it is governed by molecular interactions set by quantum effects. Hence we can suppose that there is a characteristic value of viscosity-related quantities involving $\hbar$, as in (\ref{nu0}) and (\ref{iota0}).

In addition to condensed matter, the universal lower bound of viscosity is important in high-energy physics and strongly-interacting quantum field theory. Using the duality between strongly-interacting field theories and gravity models, Kovtun, Son and Starinets (KSS) have found \cite{kss} a universal ratio between fluid viscosity and volume density of entropy $s$ as

\begin{equation}
\frac{\eta}{s}=\frac{\hbar}{4\pi k_{\rm B}}
\label{bound}
\end{equation}

This result has generated an ample interest from theoretical perspective and from the point of view of understanding the properties of quark-gluon plasma and its viscosity in particular. Relations of this result to a wider range of systems and more general effects have been of subsequent interest, including Planckian dissipation (see, e.g., Refs. \cite{zaanen1,zaanen2,hartnoll,spin1,spin2,hartnoll1} for review). KSS have conjectured that $\frac{\eta}{s}$ has a lower bound that more generally follows from strongly-coupled quantum field theories: $\frac{\eta}{s}\ge\frac{\hbar}{4\pi k_{\rm B}}$ and found that the bound is about 25 times smaller than the viscosity minima in familiar liquids such as H$_2$O and N$_2$. This raises a question of how ordinary liquids are different from high-energy hydrodynamic models. We will see that an important difference is the presence of the UV cutoff in condensed matter, setting the viscosity minima.

\section{Results and discussion}

\subsection{Kinematic viscosity}

We start with recalling the origin of viscosity minima shown in Fig. 1a where we collected available experimental $\eta$ \cite{nist} for several noble (Ar, Ne and He), molecular (H$_2$, N$_2$, CO$_2$, CH$_4$, O$_2$ and CO) and network (H$_2$O) fluids. For some fluids, we show the viscosity minimum at two pressures. The low pressure was chosen to be far above the critical pressure so that the viscosity minimum is not affected by near-critical anomalies. The highest pressure was chosen to (a) make the pressure range considered as wide as possible and (b) be low enough in order to see the viscosity minima in the temperature range available experimentally.

\begin{figure}
\begin{center}
{\scalebox{0.37}{\includegraphics{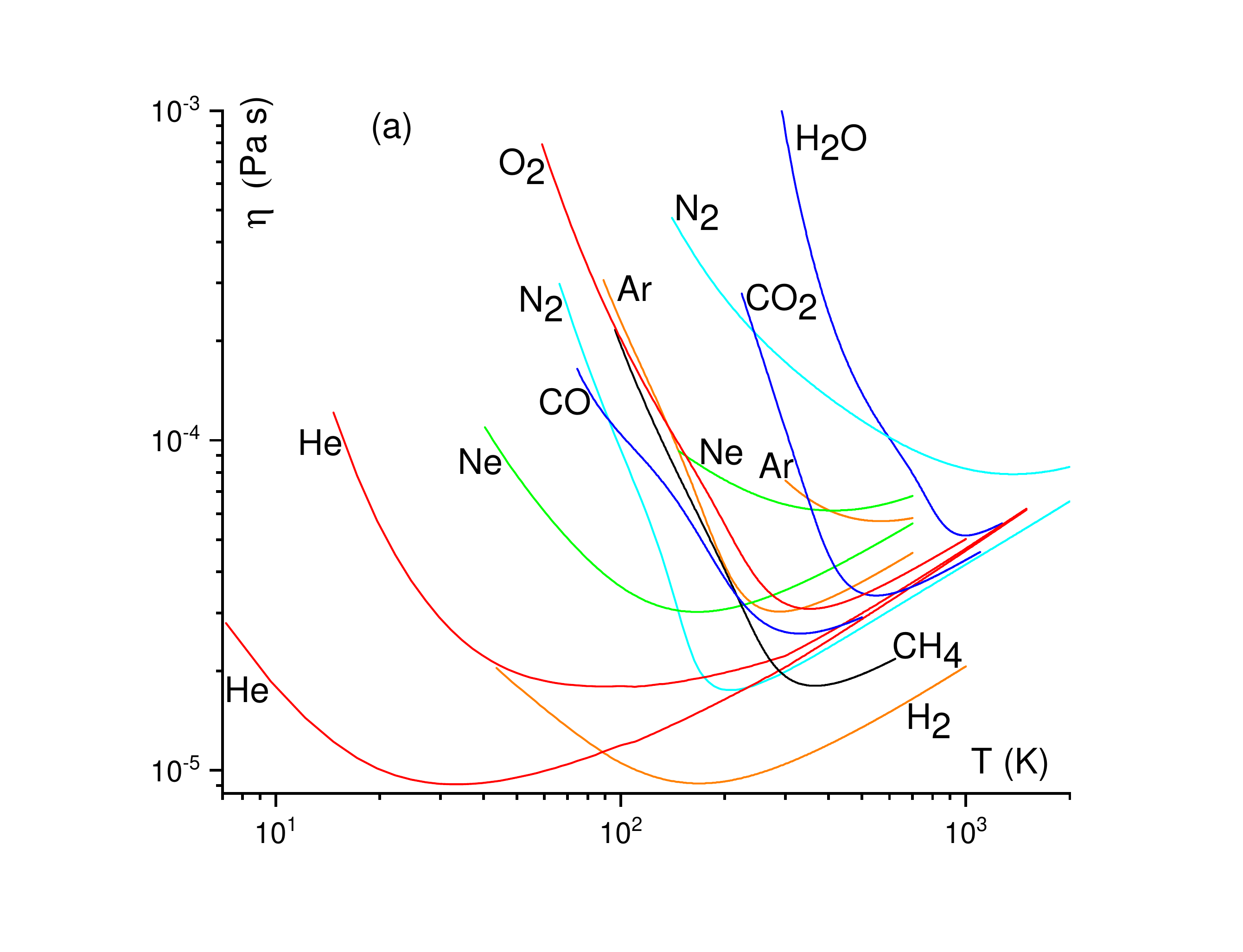}}}
{\scalebox{0.37}{\includegraphics{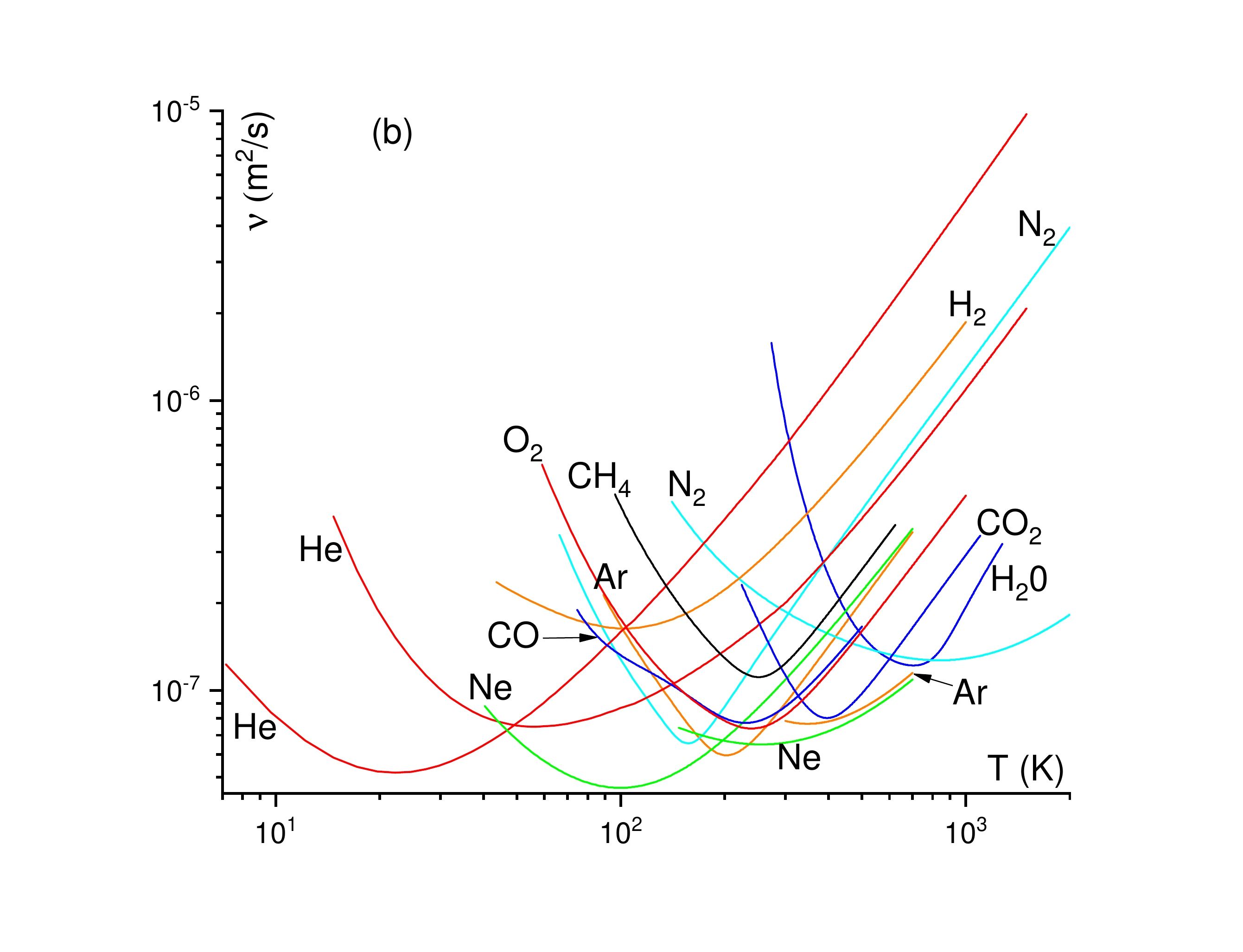}}}
\end{center}
\caption{ {\bf Viscosity and kinematic viscosity of fluids}. Experimental viscosity $\eta$ (a) and kinematic viscosity $\nu$ (b) of noble, molecular and network liquids \cite{nist} showing minima. $\eta$ for H$_2$, H$_2$O and CH$_4$ are shown for pressure $P=50$ MPa, 100 MPa and 20 MPa, respectively. $\eta$ for He, Ne, Ar and N$_2$ are shown at two pressures each: 20 and 100 MPa for He, 50 and 300 MPa for Ne, 20 and 100 MPa for Ar and 10 and 500 MPa for N$_2$. The minimum at higher pressure is above the minimum at lower pressure for each fluid.}
\label{3}
\end{figure}

In the liquid-like regime of molecular dynamics at low temperature, $\eta$ decreases with temperature as

\begin{equation}
\eta=\eta_0\exp\left(\frac{U}{T}\right)
\label{v1}
\end{equation}

\noindent where $\eta_0$ is a pre-factor and $U$ can be temperature dependent, resulting in the super-Arrhenius temperature dependence.

In the gas-like regime of molecular dynamics, $\eta$ is

\begin{equation}
\eta=\frac{1}{3}\rho v L
\label{v2}
\end{equation}

\noindent where $\rho$ is density, $v$ is average particle velocity and $L$ is the particle mean free path.

For gases, $L\propto\frac{1}{\rho}$ and $\eta\propto v\propto\sqrt{T}$ \cite{chapman}. Hence $\eta$ increases with temperature without bound, although new effects such as ionization start operating at higher temperature and change the system properties and $\eta$.

(\ref{v1}) and (\ref{v2}) imply that $\eta$ should have a minimum as seen in Fig. 1a.

Before calculating $\eta$ at the minimum, it is useful to qualify the terms ``liquid-like'' and ``gas-like'' regimes of molecular dynamics and the conditions at which the minima are seen. Molecular motion in low-temperature liquids combine solid-like oscillations around quasi-equilibrium positions and diffusive jumps to new positions, enabling liquid flow. These jumps are due to temperature-induced molecular jumps over an energy barrier set by the interaction with other molecules, resulting in (\ref{v1}). The jumps are characterised by liquid relaxation time, $\tau$, the average time between the molecular jumps, which is related to $\eta$ by the Maxwell relationship $\eta=G\tau$, where $G$ is the high-frequency shear modulus \cite{frenkel}. $\tau$ decreases with temperature in the same way as (\ref{v1}) and is bound by the elementary vibration period, commonly approximated by the Debye vibration period in the Debye model, $\tau_{\rm D}$. At this point, the oscillatory component of molecular motion is lost, and molecules start moving in a purely diffusive manner. At high temperature and/or low density, molecules gain enough energy to move distance $L$ without collisions. In this gas-like regime, the fluid viscosity can be calculated by assuming that a molecule's momentum is unchanged between collisions, resulting in (\ref{v2}).

If the temperature is increased at pressure below the critical point, the system crosses the boiling line and undergoes the liquid-gas transition. As a result, $\eta$ undergoes a sharp change at the phase transition, rather than showing a smooth minimum as in Fig. 1a. In order to avoid the effects related to the phase transition, we need to consider the supercritical state. Here, the Frenkel line \cite{f1,f2,f3} formalises the qualitative change of molecular dynamics from combined oscillatory and diffusive, where $\eta$ is given by the activation behavior (\ref{v1}), to purely diffusive, where $\eta$ follows the gas-like behavior (\ref{v2}). The location of the minima of $\eta$ slightly depends on the path taken on the phase diagram. As a result, the minimum of $\eta$ may deviate from the FL depending on the path \cite{f3}.

We now calculate viscosity at the minimum, $\eta_m$. There are two ways in which $\eta_m$ can be evaluated, by taking the low-temperature limit of the gas-like viscosity (\ref{v2}) or taking the high-temperature limit of the liquid-like viscosity given by the Maxwell relation $\eta=G\tau$. We start from the high-temperature gas-like dynamics, corresponding to the hydrodynamic regime considered in high-energy physics calculations \cite{kss} and consider how $\eta=\rho vL$ changes with temperature decrease (we drop the factor $\frac{1}{3}$ in (\ref{v2}) since our calculation evaluates the order of magnitude of viscosity minimum as discussed below). $\eta$ and the mean free path $L$ decrease on lowering the temperature but, differently from scale-invariant quantum field theories, $L$ is bound by a UV cutoff in condensed matter systems: inter-particle separation $a$ where inter-molecular interactions become appreciable (a similar effect is related to the mean free path of quasiparticles limited by the interatomic separation at the Ioffe-Regel crossover or the phonon mean free limited by the atomistic scale \cite{kittel,slack}). From this point on, $L$ has no room to decrease further. Instead, the system enters the liquid-like regime where $\eta$ starts increasing on further temperature decrease according to (\ref{v1}) because the diffusive motion of molecules crosses over to thermally-activated as discussed earlier. Therefore, the minimum of $\eta$ approximately corresponds to $L\approx a$. When $L$ becomes comparable to $a$, $v$ can be evaluated as $v=\frac{a}{\tau_{\rm D}}$ because the time for a molecule to move distance $a$ in this regime is given by the characteristic time scale set by $\tau_{\rm D}$. Setting $L=a$, $v=\frac{a}{\tau_{\rm D}}=\frac{1}{2\pi}\omega_{\rm D}a$, where $\omega_{\rm D}$ is Debye frequency, and $\rho\approx\frac{m}{a^3}$ gives $\eta_m$ as

\begin{equation}
\eta_m=\frac{1}{2\pi}\frac{m\omega_{\rm D}}{a}
\label{v3}
\end{equation}

We note that (\ref{v2}) applies in the regime where $L$ is larger than $a$, and in this sense our evaluation of viscosity minimum is an order-of-magnitude estimation, as are our other results below. In this regard, we note that theoretical models can only describe viscosity in a dilute gas limit where perturbation theory applies \cite{chapman}, but not in the regime where $L$ is comparable to $a$ and where the energy of inter-molecular interaction is comparable to the kinetic energy. In view of theoretical issues as well as many orders of magnitude by which $\eta$ can vary, we consider our evaluation meaningful. In addition to be informative, an order-of-magnitude evaluation is perhaps unavoidable if a complicated property such as viscosity is to be expressed in terms of fundamental constants only.

$\eta_m$ in (\ref{v3}) approximately matches the result obtained by approaching the viscosity minimum from low temperature where $\eta$ is given by (\ref{v1}) and considering the Maxwell relationship $\eta=G\tau$. $\eta$ and $\tau$ decrease with temperature according to (\ref{v1}), but this decrease is bound from below because $\tau$ starts approaching the shortest time scale in the system given by the Debye vibration period, $\tau_{\rm D}$. From this point on, $\tau$ has no room to decrease further. Instead, the system enters the gas-like regime where $\eta$ starts increasing with temperature according to (\ref{v2}) because the thermally-activated motion of molecules crosses over to diffusive as discussed earlier. Therefore, the minimum of $\eta$ can be approximately evaluated from $\tau\approx\tau_{\rm D}$. $G$ can be estimated as $G=\rho c^2$, where $c\approx\frac{a}{\tau_{\rm D}}$ is the transverse speed of sound. Then, $\eta_m=G\tau_{\rm D}=\rho\frac{a^2}{\tau_{\rm D}}=\frac{1}{2\pi}\frac{m\omega_{\rm D}}{a}$ as in (\ref{v3}), where we used $\rho=\frac{m}{a^3}$ as before.

Before calculating the kinematic viscosity, we first see how well (\ref{v3}) estimates the minima of $\eta$. Taking the typical values of $a=$3-6 \AA, $\frac{\omega_{\rm D}}{2\pi}$ on the order of 1 THz and atomic weights 2-40 for liquids in Fig. 1a, we find $\eta_m$ in the range $10^{-5}-10^{-4}$ Pa$\cdot$s, providing an order of magnitude estimation of $\eta_m$ consistent with Fig. 1a. We also observe that high pressure reduces $a$ and increases $\omega_{\rm D}$. Eq. (\ref{v3}) predicts that $\eta_m$ increases as a result, in agreement with the experimental behavior in Fig. 1a.

The viscosity minima of strongly-bonded metallic liquids were not measured due to their high critical points, however we note that high-temperature $\eta$ is close to $10^{-3}$ Pa$\cdot$s for Fe (2000 K), Zn (1100 K), Bi (1050 K) \cite{metals}, Hg (573 K) and Pb (1173 K) and is expected to be close to $\eta$ at the minima. This is larger than $\eta_m$ in Fig. 1a and is consistent with Eq. (\ref{v3}) predicting that $\eta_m$ increases with molecular mass ($m\omega_{\rm D}\propto\sqrt{m}$) and decreases with $a$ ($a$ is smaller in metallic systems as compared to noble and molecular ones in Fig. 1a).

As discussed above, the minimum of viscosity is ultimately related to the UV cutoff in condensed matter such as inter-particle spacing $a$ or characteristic time scale $\tau_{\rm D}$. This cutoff is absent in common scale-free field theories used in high-energy physics (see, however, Ref. \cite{matteo} where the cutoff is discussed). Below we relate the UV cutoff to fundamental physical constants.

We now consider the kinematic viscosity $\nu$ shown in Fig. 1b. $\nu$ also describes momentum diffusivity, analogous to thermal diffusivity involved in heat transfer and gives the diffusion constant in the gas-like regime of molecular dynamics \cite{frenkel}. Another benefit of considering $\nu$ is that it makes the link to the high-energy result (\ref{bound}) where $\eta$ is divided by the volume density of entropy. Using $\nu=\frac{\eta}{\rho}=vL$, $v=\frac{1}{2\pi}a\omega_{\rm D}$ and $L=a$ as before gives the minimal value of $\nu$, $\nu_m$, as

\begin{equation}
\nu_m=\frac{1}{2\pi}\omega_{\rm D}a^2
\label{nu}
\end{equation}

An expression similar to Eq. (\ref{nu}) was heuristically obtained for thermal diffusivity and interpreted as the random walk of heat transfer consisting of jumps distance $a$ with a certain frequency \cite{diff1}.

We now recall that the properties defining the UV cutoff in condensed matter can be expressed in terms of fundamental physical constants. Two quantities of interest are Bohr radius, $a_{\rm B}$, setting the characteristic scale of inter-particle separation on the order of Angstrom:

\begin{equation}
a_{\rm B}=\frac{4\pi\epsilon_0\hbar^2}{m_e e^2}
\label{bohr}
\end{equation}

\noindent and the Rydberg energy, $E_{\rm R}=\frac{e^2}{8\pi\epsilon_0a_{\rm B}}$ \cite{ashcroft}, setting the characteristic scale for the cohesive energy in condensed matter phases on the order of several eV:

\begin{equation}
E_{\rm R}=\frac{m_ee^4}{32\pi^2\epsilon_0^2\hbar^2}
\label{rydberg}
\end{equation}

\noindent where $e$ and $m_e$ are electron charge and mass.

We now recall the known ratio between the characteristic phonon energy $\hbar\omega_{\rm D}$ and the cohesive energy $E$, $\frac{\hbar\omega_{\rm D}}{E}$. Approximating $\hbar\omega_{\rm D}$ as $\hbar\left(\frac{E}{ma^2}\right)^{\frac{1}{2}}$, taking the ratio $\frac{\hbar\omega_{\rm D}}{E}$ and using $a=a_{\rm B}$ from (\ref{bohr}) and $E=E_{\rm R}$ from (\ref{rydberg}) gives, up to a factor close to 1:

\begin{equation}
\frac{\hbar\omega_{\rm D}}{E}=\left(\frac{m_e}{m}\right)^{\frac{1}{2}}
\label{ratio}
\end{equation}

We note that the factor $\left(\frac{m_e}{m}\right)^{\frac{1}{2}}$ also appears in the ratio of sound and melting velocity \cite{hartnoll1}. Combining (\ref{nu}) and (\ref{ratio}) gives

\begin{equation}
\nu_m=\frac{1}{2\pi}\frac{E a^2}{\hbar}\left(\frac{m_e}{m}\right)^{\frac{1}{2}}
\label{nu01}
\end{equation}

$a$ and $E$ in (\ref{nu01}) are set by their characteristic scales $a_{\rm B}$ and $E_{\rm R}$ as discussed earlier. Using $a=a_{\rm B}$ from (\ref{bohr}) and $E=E_{\rm R}$ from (\ref{rydberg}) in (\ref{nu01}) gives a remarkably simple $\nu_m$ as

\begin{equation}
\nu_m=\frac{1}{4\pi}\frac{\hbar}{\sqrt{m_em}}
\label{nu1}
\end{equation}

(\ref{nu1}) is one of the main results of this paper.

The same result for $\nu_m$ in (\ref{nu1}) can be obtained without explicitly using $a_{\rm B}$ and $E_{\rm R}$ in (\ref{nu01}). The cohesive energy, or the characteristic energy of electromagnetic interaction, is

\begin{equation}
E=\frac{\hbar^2}{2m_ea^2}
\label{direct}
\end{equation}

Using (\ref{direct}) in (\ref{nu01}) gives (\ref{nu1}).

Another way to derive (\ref{nu1}) is to consider the ``characteristic'' viscosity $\eta^*$ \cite{reduced}:

\begin{equation}
\eta^*=\frac{(E m)^\frac{1}{2}}{a^2}
\label{etastar}
\end{equation}

$\eta^*$ is used to describe scaling of viscosity on the phase diagram. For example, the ratio between viscosity and $\eta^*$ is the same for systems described by the same interaction potential in equivalent points of the phase diagram. For systems described by the Lennard-Jones potential, the experimental and calculated viscosity near the triple point and close to the melting line is about 3 times larger than $\eta^*$ \cite{reduced,vadim}. Near the critical point, $\eta^*$ is about 4 times larger than viscosity near the critical point and is expected to give the right order of magnitude of viscosity at the minimum at moderate pressure. The kinematic viscosity corresponding to (\ref{etastar}) is

\begin{equation}
\frac{\eta^*}{\rho}=\frac{E^\frac{1}{2}a}{m^{\frac{1}{2}}}
\label{etastar1}
\end{equation}

Using $a=a_{\rm B}$ from (\ref{bohr}) and $E=E_{\rm R}$ from (\ref{rydberg}) in (\ref{etastar1}) gives the same result as (\ref{nu1}) up to a constant factor on the order of unity. As before, we can also use (\ref{direct}) in (\ref{etastar1}) to get the same result.

We now analyze (\ref{nu1}) and its implications. $\nu_m$ contains $\hbar$ and electron and molecule masses only. $m$ characterises the molecules involved in viscous flow. $m_e$ characterises electrons setting the inter-molecular interactions.

$m$ in (\ref{nu1}) is $m=Am_p$, where $A$ is the atomic weight and $m_p$ is the proton mass. The inverse square root dependence $\nu_m\propto\frac{1}{\sqrt{A}}$ interestingly implies that for different liquids $\nu_m$ varies by a factor of about 10 only.

Setting $m=m_p$ ($A=1$) for H in (\ref{nu1}) (similarly to (\ref{bohr}) and (\ref{rydberg}) derived for the H atom) gives the fundamental kinematic viscosity in terms of $\hbar$, $m_e$ and $m_p$ as

\begin{equation}
\nu_f=\frac{1}{4\pi}\frac{\hbar}{\sqrt{m_em_p}}
\label{nuf}
\end{equation}

\noindent on the order of $10^{-7} \frac{\rm{m}^2}{\rm{s}}$.

The quantum origin of $\nu_m$, signified by $\hbar$ in (\ref{nu1}), is due to the quantum nature of inter-particle interactions. We note that in the Eyring theory, the viscosity pre-factor $\eta_0$ in (\ref{v1}) also contains $\hbar$ \cite{eyring}. This follows from assuming that the frequency of molecular oscillation in a single minimum, $\omega_0$ (attempt frequency) is set by the frequency of excited phonons as $\hbar\omega_0=k_{\rm B}T$. In later works, the pre-factor was mostly treated as a fitting parameter but its quantum nature was not examined further.


In Table 1 we compare $\nu_m$ calculated according to (\ref{nu1}) to the experimental $\nu_m$ \cite{nist} for all liquids shown in Fig. 1. The ratio between experimental and predicted $\nu_m$ is in the range of about 0.5-3. For the lightest liquid, H$_2$, experimental $\nu_m$ is close to the theoretical fundamental viscosity (\ref{nuf}). We therefore find that (\ref{nu1}) predicts the right order of magnitude of $\nu_m$.

\begin{table}[ht]
\begin{tabular}{ l l l}
\hline
                   & $\nu_m$ (calc.)       & $\nu_m$ (exp.)\\
                   & $\times$10$^8$ m$^2$/s & $\times$10$^8$ m$^2$/s\\
\hline
Ar (20 MPa)        & 3.4        & 5.9 \\
Ar (100 MPa)       & 3.4        & 7.7 \\
Ne (50 MPa)        & 4.8        & 4.6 \\
Ne (300 MPa)       & 4.8        & 6.5 \\
He (20 MPa)        & 10.7       & 5.2 \\
He (100 MPa)       & 10.7       & 7.5 \\
N$_2$ (10 MPa)     & 4.1        & 6.5 \\
N$_2$ (500 MPa)    & 4.1        & 12.7 \\
H$_2$ (50 MPa)     & 15.2       & 16.3 \\
O$_2$ (30 MPa)     & 3.8        & 7.4 \\
H$_2$O (100 MPa)   & 5.1        & 12.1 \\
CO$_2$ (30 MPa)    & 3.2        & 8.0 \\
CH$_4$ (20 MPa)    & 5.4        & 11.0 \\
CO (30 MPa)        & 4.1        & 7.7 \\
\hline \label{table}
\end{tabular}
\caption{Calculated and experimental $\nu_m$.}
\end{table}

We observe that $\nu_m$ increases with pressure in Table 1, similarly to $\eta_m$ in Fig. 1. However, pressure dependence is not accounted in $\nu_m$ in (\ref{nu1}) since (\ref{nu1}) is derived in the approximation involving Eqs. (\ref{bohr})-(\ref{nu01}) which do not account for the pressure dependence of $\omega_{\rm D}$ and $E$.

We make four further remarks regarding the comparison in Table 1. First, the important term in Eq. (\ref{nu1}) is the combination of fundamental constants $\frac{\hbar}{\sqrt{m_em}}$ which sets the characteristic scale of the minimal kinematic viscosity, whereas the numerical factor in (\ref{nu1}) may be affected by the approximations used and mentioned earlier. Second, Eqs. (\ref{bohr})-(\ref{ratio}) assume valence electrons directly involved in bonding and hence strongly-bonded systems, including metallic, covalent and ionic liquids. Their viscosity in the supercritical state is unavailable due to high critical points. The available data in Fig. 1 and Table 1 includes weakly-bonded systems such as noble, molecular and hydrogen-bonded fluids. Although bonding in these systems is also electromagnetic in origin, weak dipole and van der Waals interactions result in smaller $E$ and, consequently, smaller $\eta$ as compared to strongly-bonded ones, with the viscosity of hydrogen-bonded fluids lying in between \cite{vadim1}. However, $\nu_m$ in (\ref{nu01}) and $\eta$ in (\ref{etastar1}) contain factors $Ea^2$ and $E^{\frac{1}{2}}a$, respectively. $E^{\frac{1}{2}}$ is 3-10 times smaller and $a$ is 2-4 times larger in weakly-bonded as compared to strongly-bonded systems \cite{vadim1}. This results in a weak dependence of $\nu_m$ on bonding type, and the order-of-magnitude evaluation (\ref{nu1}) is unaffected as Table 1 shows. Third, Eq. (\ref{nu1}) for strongly-bonded fluids serves as a {\it prediction} for future experimental work.

Fourth and finally, the above argument regarding the minimum of $\nu_m$ in (\ref{nu1}) corresponds to the liquid and supercritical states. However,

\subsection{Elementary viscosity}

Eq. (\ref{nuf}) gives the maximal value of the minimum of kinematic viscosity for H. It is interesting to ask what viscosity-related quantity has an absolute minimum. We introduce a new quantity: the ``elementary'' viscosity $\iota$ (``iota'') defined as the product of $\eta_m$ and elementary volume $a^3$: $\iota=\eta_m a^3$ or, equivalently, as $\iota=\nu_m m$. Using (\ref{nu1}), $\iota$ is

\begin{equation}
\iota=\frac{\hbar}{4\pi}\left({\frac{m}{m_e}}\right)^{\frac{1}{2}}
\label{iota1}
\end{equation}

\noindent which has the lower bound, $\iota_m$, for $m=m_p$ in H:

\begin{equation}
\iota_m=\frac{\hbar}{4\pi}\left({\frac{m_p}{m_e}}\right)^{\frac{1}{2}}
\label{iota2}
\end{equation}

\noindent and is on the order of $\hbar$.

\begin{figure}
\begin{center}
{\scalebox{0.35}{\includegraphics{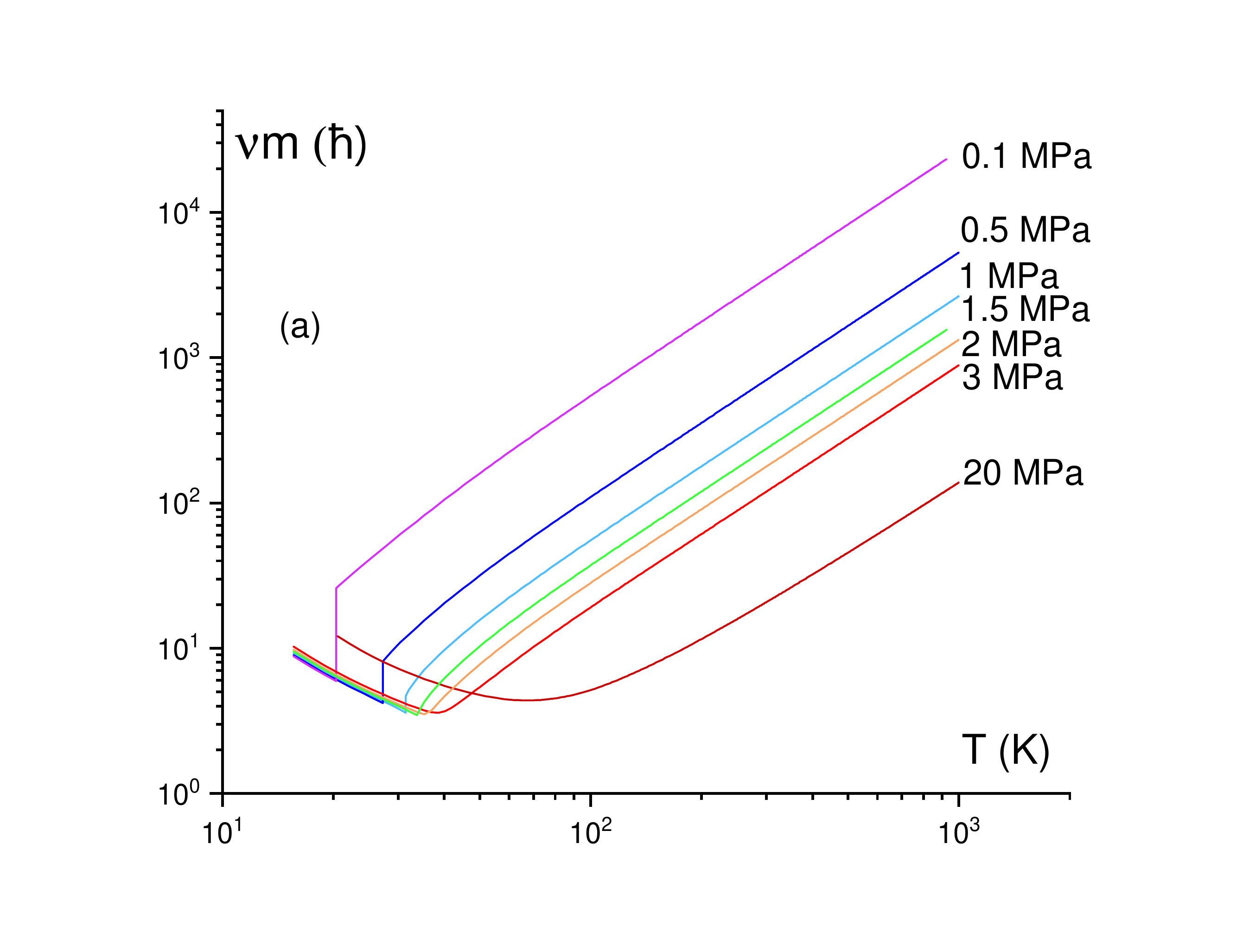}}}
{\scalebox{0.35}{\includegraphics{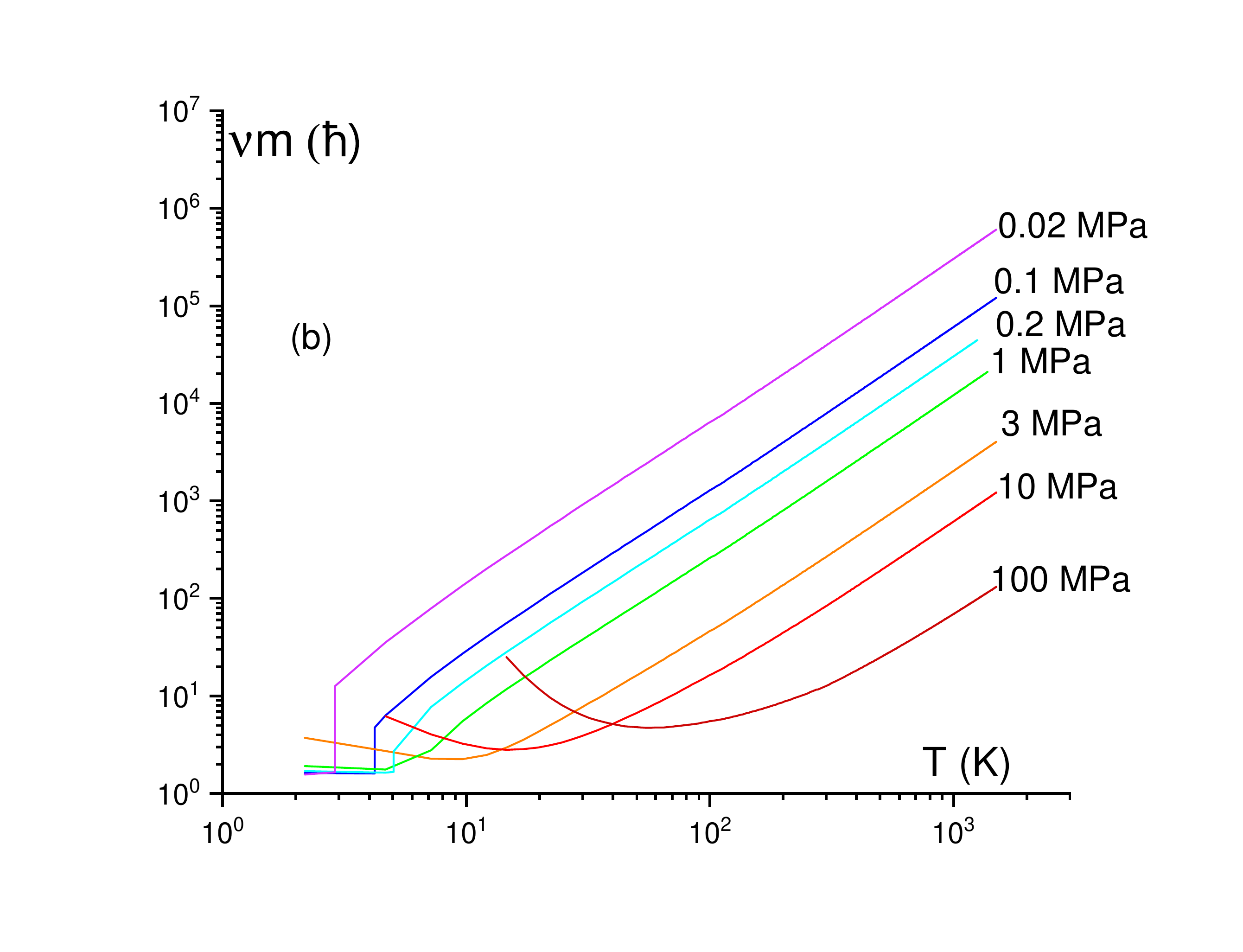}}}
{\scalebox{0.35}{\includegraphics{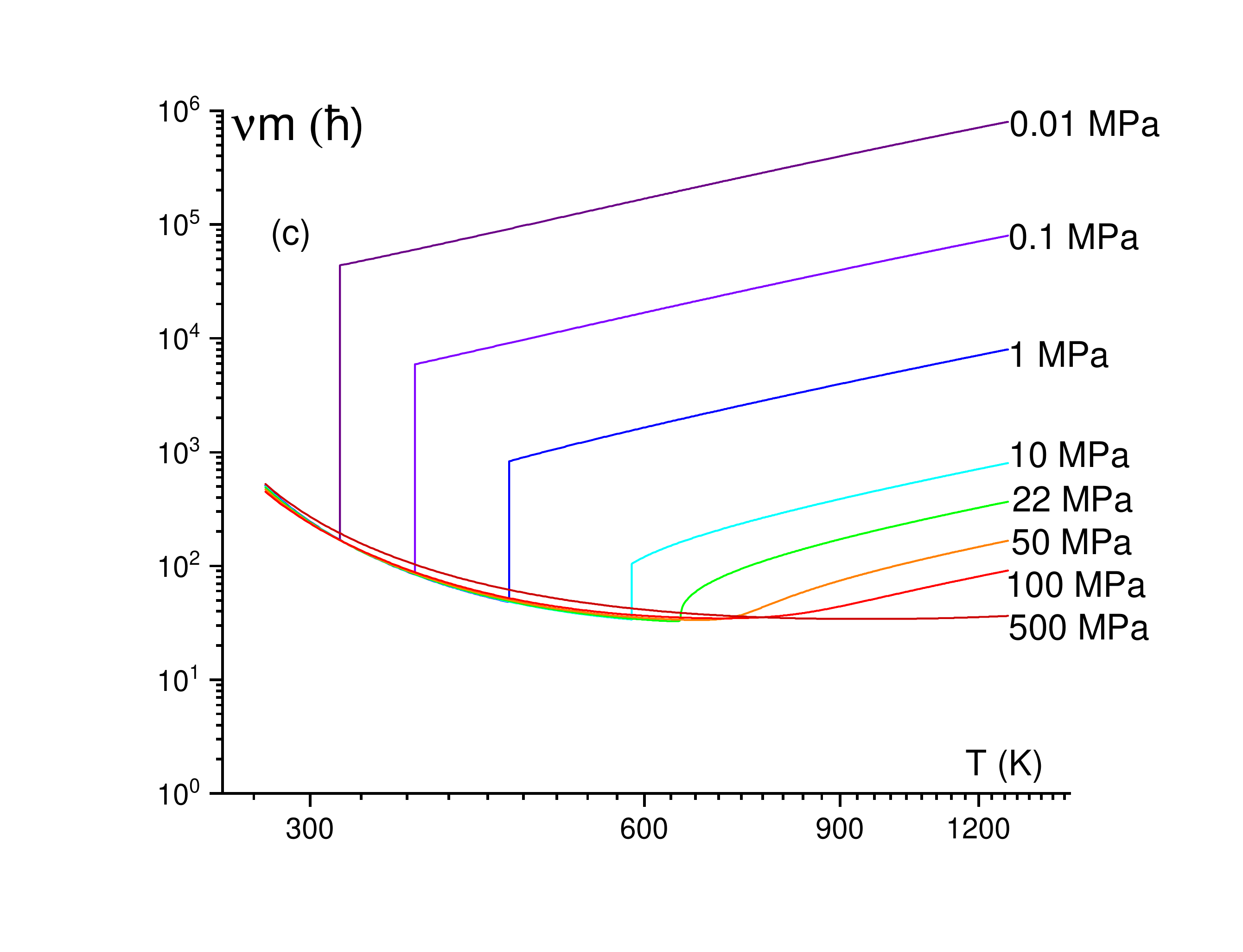}}}
\end{center}
\caption{{\bf Elementary viscosity of fluids}. $\nu m$ calculated from experimental kinematic viscosity \cite{nist} for H$_2$ (a), He (b) and H$_2$O (c) below and above the critical pressure $P_c$. $P_c$=1.3 MPa for H$_2$, 0.23 MPa for He and 22 MPa for H$_2$O. The smallest value of $\nu m$, $\iota$, is consistent the lower bound (\ref{iota2}).}
\label{3}
\end{figure}

(\ref{iota2}) notably involves the proton-to-electron mass ratio, one of few dimensionless combinations of fundamental constants of general importance \cite{barrow}.

In Fig. 2a-b, we show the product $\nu m$ in the units of $\hbar$ for two lightest liquids, H$_2$ and He, for which the minimum of $\nu m$, $\nu_m m=\iota$, should be the closest to the lower bound (\ref{iota2}). We calculate $\nu m$ using the experimental data \cite{nist} and show it above and below the critical pressure $P_c$. For He, the temperature range is above the superfluid transition (we do not consider superfluidity in this work).

The liquid-gas phase transition results in sharp changes of viscosity below $P_c$. For H$_2$, the minimum of $\nu m$ is kinked as a result and decreases with pressure up to $P_c$. This is followed by the minimum becoming smooth and increasing above $P_c$. We observe that the smooth minimum just above the critical point (where our derivation of $\eta_m$ and $\nu_m$, assuming a non-interrupted variation of viscosity, applies) is very close to the minimum at $P_c$. For He, the minimum similarly increases with pressure in the supercritical region and weakly varies below $P_c$.

The smallest value of $\nu m$, $\iota=\nu_m m$, in Fig. 2a-b is in the range (1.5-3.5)$\hbar$ for He and H$_2$. This is consistent with the estimation of the lower bound of $\iota$, $\iota_m$ in (\ref{iota2}). Given that $\nu m$ varies 4-6 orders of magnitude in Fig. 2, the agreement with our result (\ref{iota2}) is notable.

We also show $\nu m$ for common H$_2$O in Fig. 2c as a reference and include the triple and critical point in the pressure range. The qualitative behavior of $\nu m$ is similar to that of H$_2$, with $\iota$ of about 30$\hbar$.

Our lower bound for $\iota$ is consistent with the uncertainty principle. As discussed earlier, the minimum of $\nu$ can be evaluated as $\nu_m=va$, corresponding to $\iota=mva=pa$, where $p$ is particle momentum. According to the uncertainty principle applied to a particle localised in the region set by $a$, $\iota\ge\hbar$. This is consistent with our bound $\iota_m$ in (\ref{iota2}), although a more general (\ref{iota1}) gives a stronger bound which increases for heavier molecules.

We note that the data showing the increase of $\nu$ with temperature in the gas regime in Fig. 1 and 2 is above the triple point. Below the triple point, $\nu$ is larger than our lower bound. This follows from observing that (a) $\nu$ increases above the sublimation line along the isobars and also increases along the sublimation line on lowering the temperature due to the exponential decrease of sublimation pressure \cite{landau} (we do not consider quantum effects), and (b) $\nu$ at the triple point is significantly larger than our minimum \cite{nist}. Therefore, $\nu_m$ and $\iota_m$ correspond to the minimum for both fluids and gases (phases where viscosity operates), including dilute low-temperature gases.

We now return to the high-energy physics result and the finding of Kovtun, Son and Starinets \cite{kss} that their lower bound (\ref{bound}) is about 25 times smaller than in liquid H$_2$O and N$_2$. We consider, more generally, the ratio between $\eta$ and the volume density $d=\frac{qN}{V}$ of any intensive quantity $Q=qN$, where $N$ is the number of particles. Then, $\frac{\eta}{d}=\frac{\nu m}{q}$. If $Q$ is entropy $S=qN$ and $q_0$ is $q$ corresponding to $\nu_m$, the experimental data \cite{nist} show that the minimum of $\frac{\eta}{d}$ or $\frac{\nu m}{q}$, $R_m$, is close to $\frac{\nu_m m}{q_0}=\frac{\iota}{q_0}$ due to slow temperature variation of entropy. Then, $R_m$ is conveniently written in terms of $\iota$ using (\ref{iota1}) as

\begin{equation}
R_m=\frac{\iota}{q_0}=\frac{\hbar}{4\pi q_0}\left({\frac{m}{m_e}}\right)^{\frac{1}{2}}
\label{bound1}
\end{equation}

We observe that, compared to the KSS bound (\ref{bound}), (\ref{bound1}) contains an additional factor $\left({\frac{m}{m_e}}\right)^{\frac{1}{2}}$. For H$_2$O and N$_2$ considered by KSS, $\left({\frac{m}{m_e}}\right)^{\frac{1}{2}}$ is 182 and 227. Using $q_0=8.7k_{\rm B}$ for H$_2$O and $14.6k_{\rm B}$ for N$_2$ \cite{nist} at pressures considered \cite{kss}, (\ref{bound1}) predicts that the ratio $\frac{\eta}{s}$ in these liquids exceeds the bound (\ref{bound}) by a factor of 16-21, in order-of magnitude agreement with the KSS finding and the ability of (\ref{nu1}) to predict experimental $\nu_m$ within an order of magnitude.


\section{Conclusions}

In summary, we have found a new quantum quantity corresponding to the minimum of kinematic viscosity, an interesting result in view of wide variation of viscosity across different systems and external parameters as well as complexity of theoretical description. A related result is the new ``elementary'' viscosity $\iota$ with the lower bound set by fundamental physical constants and involving the proton-to-electron mass ratio.

Acknowledgements: we are grateful to M. Baggioli, A. Starinets and D. Vegh for discussions and EPSRC for support. 



\end{document}